\documentclass[fleqn,12pt]{article}
\usepackage[numbers,sort&compress]{natbib}
\usepackage[footnotesep=0.4in]{geometry}
\usepackage{graphicx}
\usepackage{caption2}
\usepackage{amsmath}
\usepackage{bbding}
\usepackage{graphics}
\usepackage{pifont}
\usepackage{amsfonts}
\usepackage{caption2}
\usepackage{float}
\makeatletter
\newcommand\figcaption{\def\@captype{figure}\caption}
\newcommand\tabcaption{\def\@captype{table}\caption}
\begin{document}

\date{}
\title{Spacial inhomogeneity and nonlinear tunneling for the forced KdV equation}
\author{  Xin Yu$^{1,}$\thanks{Corresponding
author, with e-mail address as yuxin@buaa.edu.cn}\,,\, Zhi-Yuan
Sun$^{1}$\,,\, Kai-Wen Zhou$^{1}$\,,\, Yu-Jia Shen$^{2}$
\\
\\{\em 1. Ministry-of-Education Key Laboratory of Fluid Mechanics and
National}\\
{\em  Laboratory for Computational Fluid Dynamics,  Beijing University of }\\
{\em  Aeronautics and Astronautics, Beijing 100191, China} \\
 {\em 2. Key Laboratory of Mathematics Mechanization, Institute of Systems Science,}\\
{\em    AMSS, Chinese Academy of Sciences, Beijing 100190, China}}
\maketitle

\vspace{-3mm}

\begin{abstract}

A variable-coefficient forced Korteweg-de Vries equation with
spacial inhomogeneity is investigated in this paper. Under
constraints, this equation is transformed into its bilinear form,
and multi-soliton solutions are derived.  Effects of spacial
inhomogeneity for soliton velocity, width and background are
discussed. Nonlinear tunneling for this equation is presented, where
the soliton amplitude can be amplified or compressed. Our results
might be useful for the relevant problems in fluids and plasmas.

\end{abstract}
\vspace{3mm}


\vspace{3mm}


\vspace{20mm}

\newpage
\noindent {\Large{\bf I. Introduction}}

\vspace{3mm}In this paper,  we will investigate the following
 variable-coefficient forced KdV
equation~\cite{hd,tkg,hd3,gxh,hhd,ph,hd2,rhjg,gae,tangxiaoyan} with
the aid of symbolic computation~\cite{VCKPt,VCKPg,Pengchong Jin},
\begin{equation}\label{equation}
\hspace{0mm}u_t\,+a(t)\,u\,u_{x}+b(t)\,u_{xxx}+c(t)\,u+d(t)\,u_{x}=f(x,t)\,,
\end{equation}
where $u$ is a function of the scaled spacial coordinate $x$ and
temporal coordinate $t$, $a(t)$, $b(t)$, $c(t)$ and $d(t)$ are the
analytic functions of $t$, and $f(x,t)$ is the analytic function of
$x$ and $t$. These variable coefficients respectively represent the
nonlinear, dispersive, line-damping, dissipative and external-force
effects,  which are caused by  the inhomogeneities of media and
boundaries~\cite{hd,tkg,hd3,gxh,hhd,ph,hd2,rhjg,gae,tangxiaoyan}.
Here, we assume that  the spacial inhomogeneity is linear and take
the following form
\begin{equation}\label{cc}
\hspace{10mm}f(x,t)=f_{1}(t)x+f_{2}(t)\,.
\end{equation}

When coefficients are taken with different cases,
Eq.~(\ref{equation}) has been seen to describe nonlinear waves in a
fluid-filled tube~\cite{hd,tkg,hd3}, weakly nonlinear waves in the
water of variable depth~\cite{ph,hd2}, trapped quasi-one-dimensional
Bose-Einstein condensates~\cite{gxh}, internal gravity waves in
lakes with changing cross sections~\cite{rhjg}, the formation of a
trailing shelf behind a slowly-varying solitary wave~\cite{gae},
dynamics of a circular rod composed of a general compressible
hyperelastic material with the variable cross-sections and material
density~\cite{hhd}, and atmospheric and oceanic dynamical
systems~\cite{tangxiaoyan}.

If the the spacial inhomogeneity is ignored, i.e., $f_{1}(t)=0$,
Eq.~(\ref{equation}) has been transformed into its several KdV-typed
ones with simpler forms~\cite{zhang1,tiankdv,lijuan,salas}, and  has
also been solved directly via the bilinear method~\cite{yu1}. The
effects of the dispersive, line-damping, dissipative, and
external-force terms on the solitonic velocity, amplitude and
background have been discussed~\cite{yu1} with the
characteristic-line method~\cite{characteristic line,yu}; Besides,
Wronskian form are derived based on the given bilinear
expression~\cite{yu2}.

However, since the spacial inhomogeneity in  external-force term
brings into more difficulties in solving, to our knowledge, the
multi-soliton solutions for Eq.~(\ref{equation}) in the explicit
bilinear forms have not been constructed directly, and the effects
of spacial inhomogeneity on solitonic propagation and interaction
have not been discussed.

In addition, nonlinear tunneling for the nonautonomous nonlinear
Schr\"{o}dinger equations has attracted attention in recent
years~\cite{nt1,nt2,nt3,nt4}. The concept of the nonlinear tunneling
effect comes from the wave equations steming from the nonlinear
dispersion relation, which has shown that the soliton can pass
lossless through the barrier/well under special conditions which
depend on the ratio between the amplitude of the soliton and the
height of the barrier/well~\cite{nt1,nt2,nt3,nt4}. In this paper, we
will apply such concept to Eq.~(\ref{equation}), a KdV-typed
equation. In section II, a dependent variable transformation and two
constraints will be proposed, Eq.~(\ref{equation}) will be
transformed into its bilinear form, and the multi-soliton solutions
in the explicit forms will be constructed. In section III, we will
show that different from Ref.~\cite{yu1}, the nonlinear coefficient
can  also affect the soliton width and amplitude for the existence
of spacial inhomogeneity in the forced term. In section IV, we will
discuss nonlinear barrier/well of Eq.~(\ref{equation}). Finally,
Section V will present the conclusions.

\vspace{7mm} \noindent {\Large{\bf II. Soliton solutions}}

\vspace{3mm}

Through the dependent variable transformation
\begin{equation}\label{bianhuan}
\hspace{10mm}u\,=\alpha(t)({\rm log}
\Phi)_{xx}+\beta(t)+\gamma(t)x\,,\\
\end{equation}
and  the coefficient constraints,
\begin{equation}\label{cc1}
\hspace{10mm}b(t)=\frac{\rho\,a(t)}{6}\,e^{\int{[a(t)\gamma(t)-c(t)]dt}}\,,
\end{equation}
\begin{equation}\label{cc2}
\hspace{10mm}f_{1}(t)=a(t)\gamma(t)^{2}+c(t)\gamma(t)+\gamma^{'}(t)\,,
\end{equation}
where
\begin{equation}
\hspace{10mm}\alpha(t)=2\,\rho\,e^{\int{[a(t)\gamma(t)-c(t)]dt}}\,,
\end{equation}
\begin{equation}
\hspace{10mm}\beta(t)=e^{\int{[-a(t)\gamma(t)-c(t)]dt}} \Big\{\delta
+\int{e^{\int{[a(t)\gamma(t)+c(t)]dt}}\Big[f_{2}(t)-d(t)\gamma(t)\Big]dt}\Big\}\,,
\end{equation}
 $\Phi$ is a  function of $x$ and $t$, $\rho$ and $\delta$
are constants, and $'$ denotes the derivative with respect to $t$,
Eq.~(\ref{equation}) can be transformed into the following bilinear
form,
\begin{equation}\label{bilinear}
\hspace{10mm}\Big\{D_xD_t+b(t)\,D_{x}^{4}+\Big[d(t)+a(t)\beta(t)+a(t)\gamma(t)x\Big]\,D_x^{2}
+a(t)\gamma(t)\frac{\partial}{\partial x}\Big\} \Phi \cdot \Phi=0\,,
\end{equation}
where $D_{x}^{m}D_{t}^{n}$ is the bilinear derivative
operator~\cite{hirota1,hirota2} defined by
\begin{eqnarray}
\hspace{10mm}D_{x}^{m}D_{t}^{n} a\cdot
b\equiv\left(\frac{\partial}{\partial x}-\frac{\partial}{\partial
x^{'}}\right)^{m}\,\left(\frac{\partial}{\partial
t}-\frac{\partial}{\partial
t^{'}}\right)^{n}\,a(x,t)\,b(x^{'},t^{'}) \bigg
|_{x^{'}=x,\,t^{'}=t}\,,\
\end{eqnarray}
and
\begin{equation}
\hspace{10mm}\frac{\partial}{\partial x} \Phi \cdot \Phi=2 \Phi
\Phi_x\,.
\end{equation}

Note that the independence of $f_{1}(t)$ is transformed to that of
$\gamma(t)$ through constraint~(\ref{cc2}).

We expand $\Phi$ in the power series of a  parameter $\epsilon$ as
\begin{equation}\label{exp}
\hspace{10mm}\Phi =
1\,+\,\epsilon\,\Phi_1\,+\,\epsilon^{2}\,\Phi_2\,+\,\cdots\,.\\
\end{equation}
Substituting Expansion~(\ref{exp}) into Eq.~(\ref{bilinear}) and
collecting the coefficients of the same power of $\epsilon$, through
the standard process of the Hirota bilinear method, we can derive
the $N$-soliton-like solutions for Eq.~(\ref{equation}), which can
be denoted as
\begin{equation}\label{u}
\hspace{-3mm}u=\alpha(t)\,\frac{\partial^2}{\partial
x^2}\Big\{\log{\Big[\sum_{\mu=0,1}\mathrm{exp}\Big(\sum_{j=1}^{N}\mu_j\,\xi_j+\sum_{1\leq
j<l}^{N}\mu_j\,\mu_l\,A_{j\,l}\Big)\Big]}\Big\}+\beta(t)+\gamma(t)x\,,
\end{equation}
with
\begin{equation}\label{xiang}
\hspace{10mm}\xi_j=k_j(t)\,x+\omega_j(t)+\xi_j^{0}\,,\\
\end{equation}
\begin{equation}\label{xiang2}
\hspace{10mm}k_j(t)=e^{-\int{a(t)\gamma(t)dt}}k_j\,,\\
\end{equation}
\begin{equation}\label{sesan}
\hspace{10mm}\omega_j(t)=-\int{k_j^3(t)b(t)dt}-\int{k_j(t)\Big[d(t)+a(t)\beta(t)\Big]dt}\,,\\
\end{equation}
\begin{equation}\label{xiangyi}
\hspace{10mm}e^{A_{jl}}=\frac{(k_j-k_l)^{2}}{(k_j+k_l)^{2}}\,,
\end{equation}
where $k_j$ and $\xi_j^{0}$ $(j=1,2,\cdots,N)$ are arbitrary real
constants, $\sum_{\mu=0,1}$ is a summation over all possible
combinations of $\mu_1=0,1,\,\mu_2=0,1,\,\cdots,\,\mu_N=0,1$, and
$\sum_{1\leq j<l}^{N}$ means a summation over all possible pairs
$(j,l)$ chosen from the set $(1,2,\ldots,N)$, with the condition
that $1\leq j<l$~\cite{hirota2}.

Specially, one  soliton solution can be expressed as
\begin{equation}\label{1soliton}
\hspace{10mm} \Phi=1+ \mathrm{exp}\Big[
k_1(t)\,x+\omega_1(t)+\xi_{\text{10}}\Big]\,,
\end{equation}
and two soliton solution can be expressed as
\begin{align}\label{2soliton}
&\hspace{10mm}\notag \Phi=1+ \mathrm{exp}\Big[
k_1(t)\,x+\omega_1(t)+\xi_{\text{10}}\Big] + \mathrm{exp}\Big[
k_2(t)\,x+\omega_2(t)+\xi_{\text{20}}\Big] \notag\\&\hspace{11mm}+
\mathrm{exp}\Big[ k_1(t)\,x+\omega_1(t)+\xi_{\text{10}}
+k_2(t)\,x+\omega_2(t)+\xi_{\text{20}}+A_{12}\Big]\,.
\end{align}

\vspace{7mm} \noindent {\Large{\bf III. Spacial inhomogeneity}}

\vspace{3mm}

The coefficients $a(t)$, $b(t)$, $c(t)$, $d(t)$ and $f_2(t)$ have
the similar influences on the soliton velocity, amplitude and
background, which have been discussed in Refs.~\cite{yu1,yu2}. Thus,
we will mainly discuss the influence of the spacial inhomogeneity in
the forced term.

As shown in Fig.~\ref{tu1},  the soliton width broadens, amplitude
increases, and position of soliton raises. In
expression~(\ref{xiang2}), $a(t)$ and $\gamma(t)$ occur
simultaneously, so nonlinear coefficient $a(t)$ can  affect the
soliton width and amplitude.
\\[\intextsep]
\begin{minipage}{\textwidth}
\renewcommand{\captionfont}{\scriptsize}
\renewcommand{\captionlabelfont}{\scriptsize}
\renewcommand{\captionlabeldelim}{.\,}
\renewcommand{\figurename}{Fig.\,}
\hspace{2.3cm}\includegraphics[scale=0.55]{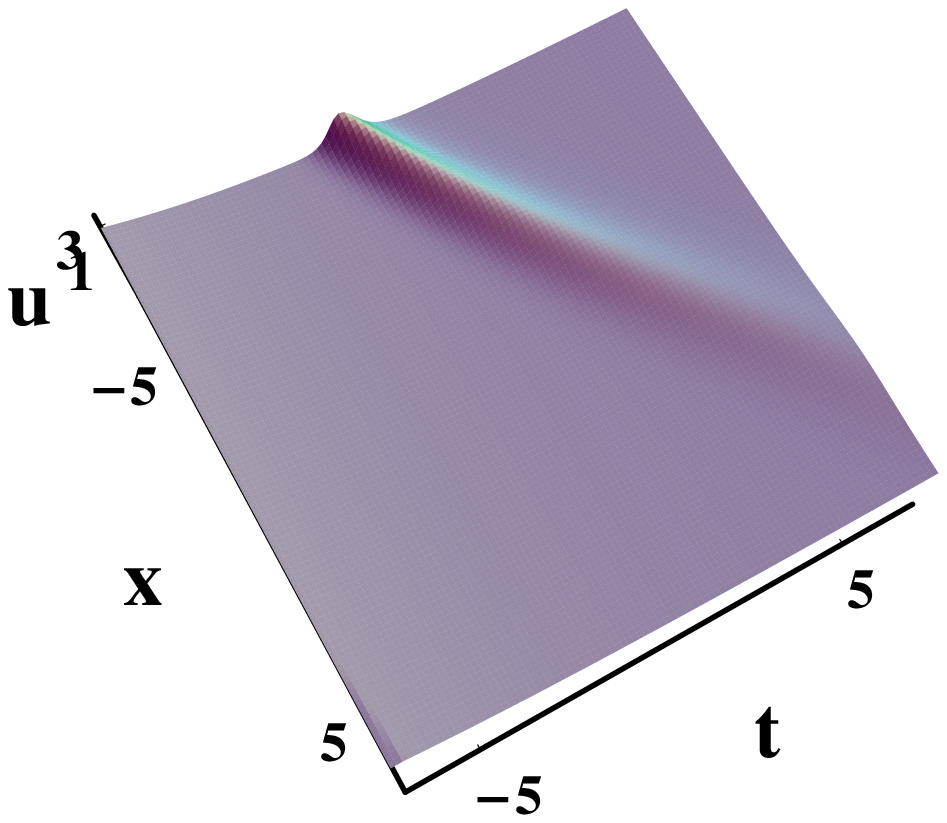}
\hspace{1.5cm}\includegraphics[scale=0.55]{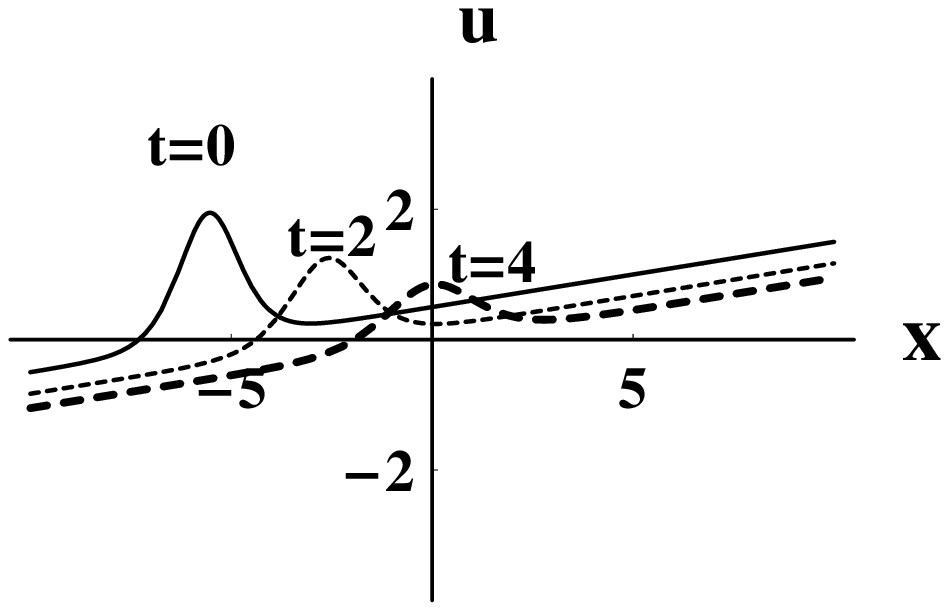}
\vspace{-0.5cm}{\center\hspace{4.7cm}\footnotesize ($a$)
\hspace{6.6cm}($b$)} \figcaption{One soliton given by
Expression~(\ref{1soliton}) with parameters: $k_{\text{1}}=2$,
$\rho=\delta=1$, $d(t)=a(t)=1$, $c(t)=\gamma(t)=0.1$,
$\xi_{\text{10}}=-10$; ($b$) Profile of Fig.1 (a) at $t=0$, $t=4$,
$t=6$.} \label{tu1}
\end{minipage}
\\[\intextsep]

Fig.~\ref{tu2} presents a case that the soliton velocity, amplitude
and background are periodic. Fig.~\ref{tu3} corresponds to the
periodic two soliton solution.
\\[\intextsep]
\begin{minipage}{\textwidth}
\renewcommand{\captionfont}{\scriptsize}
\renewcommand{\captionlabelfont}{\scriptsize}
\renewcommand{\captionlabeldelim}{.\,}
\renewcommand{\figurename}{Fig.\,}
\hspace{2.3cm}\includegraphics[scale=0.55]{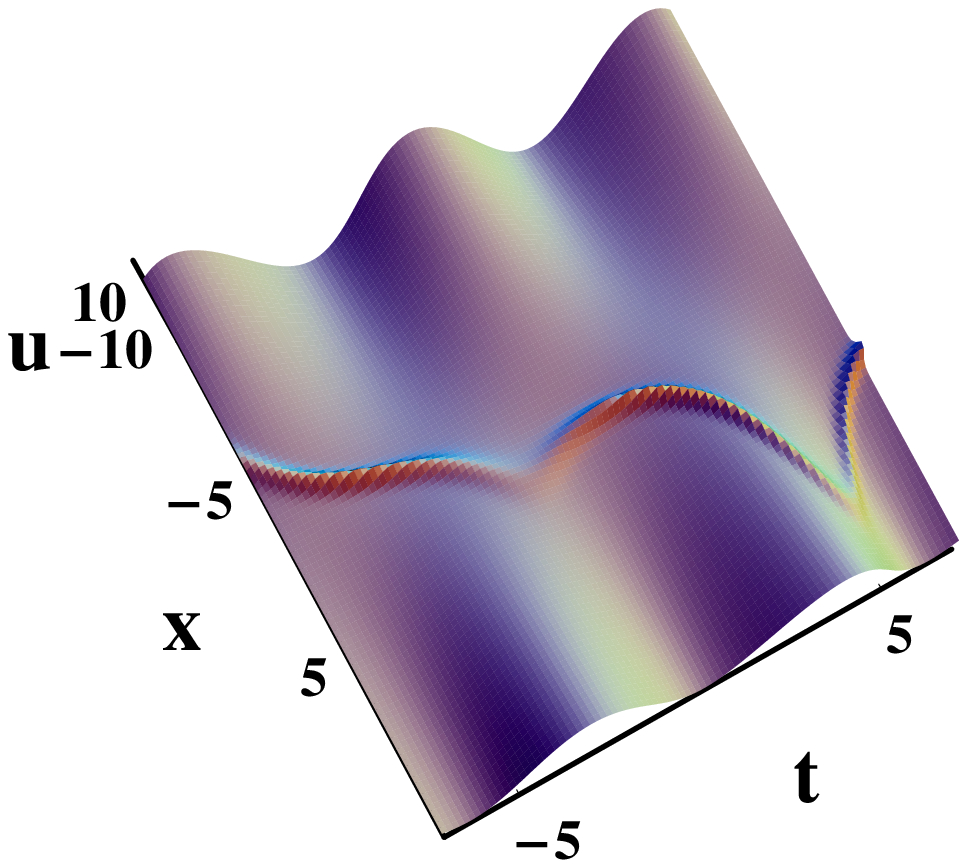}
\hspace{1.5cm}\includegraphics[scale=0.55]{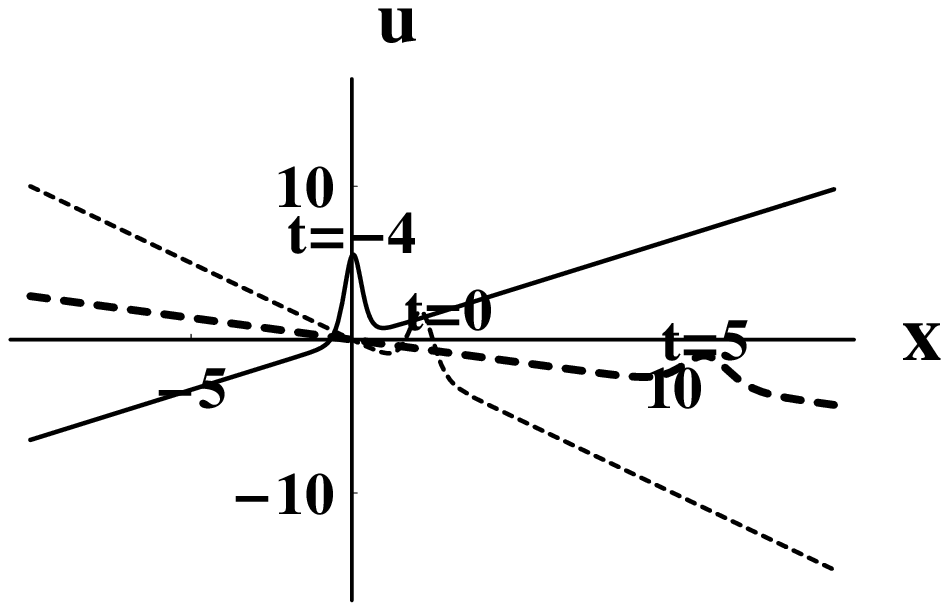}
\vspace{-0.5cm}{\center\hspace{4.7cm}\footnotesize ($a$)
\hspace{6.6cm}($b$)} \figcaption{One soliton given by
Expression~(\ref{1soliton}) with parameters: $k_{\text{1}}=2$,
$\rho=1$, $d(t)=f_{2}(t)=\delta=0$, $\xi_{\text{10}}=-10$,
$a(t)=[2+\sin(t)]^{-1}$, $e^{\int{[-a(t)\gamma(t)]dt}}=2+\sin(t)$,
$e^{\int{-c(t)dt}}=1$; ($b$) Profile of Fig.2 (a) at $t=-4$, $t=0$,
$t=5$.} \label{tu2}
\end{minipage}
\\[\intextsep]
\\[\intextsep]
\begin{minipage}{\textwidth}
\renewcommand{\captionfont}{\scriptsize}
\renewcommand{\captionlabelfont}{\scriptsize}
\renewcommand{\captionlabeldelim}{.\,}
\renewcommand{\figurename}{Fig.\,}
\hspace{2.3cm}\includegraphics[scale=0.55]{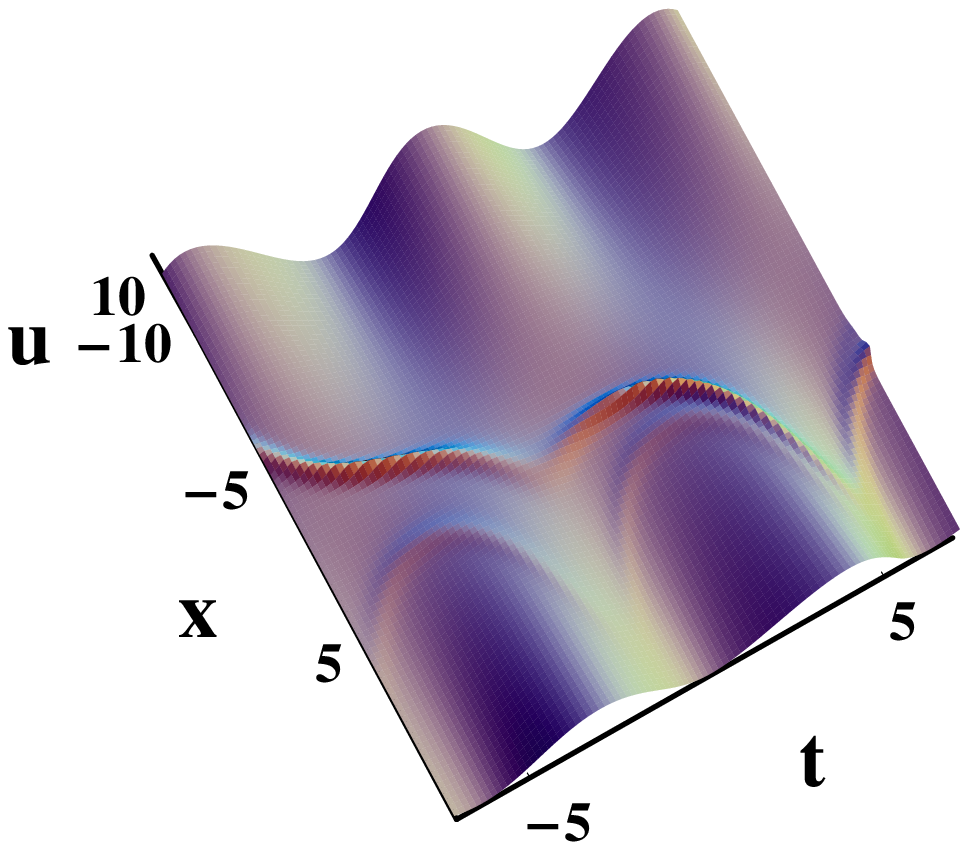}
\hspace{1.5cm}\includegraphics[scale=0.55]{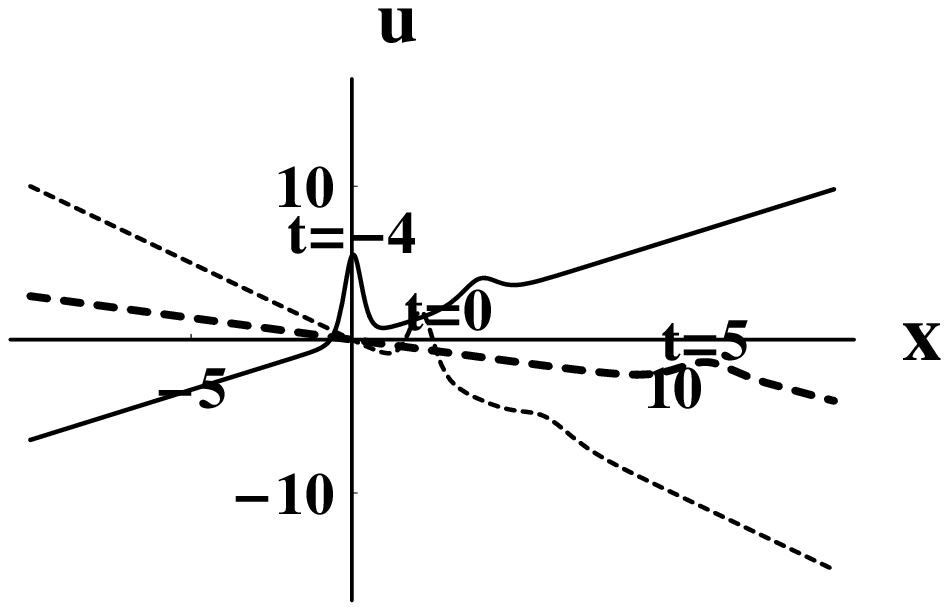}
\vspace{-0.5cm}{\center\hspace{4.7cm}\footnotesize ($a$)
\hspace{6.6cm}($b$)} \figcaption{Two  solitons given by
Expression~(\ref{2soliton}) with parameters: $k_{\text{1}}=2$,
$k_{\text{2}}=1$, $\rho=1$, $d(t)=f_{2}(t)=\delta=0$,
$\xi_{\text{10}}=-10$, $a(t)=[2+\sin(t)]^{-1}$,
$e^{\int{[-a(t)\gamma(t)]dt}}=2+\sin(t)$, $e^{\int{-c(t)dt}}=1$;
($b$) Profile of Fig.3 (a)  at $t=-4$, $t=0$, $t=5$.} \label{tu3}
\end{minipage}
\\[\intextsep]

\vspace{7mm} \noindent {\Large{\bf IV. Nonlinear tunneling}}

\vspace{3mm}

Nonlinear tunneling has been discussed  for the nonlinear
Schr\"{o}dinger equation~\cite{nt1,nt2,nt3,nt4}. Hereby, we will
investigate the nonlinear tunneling for the KdV equation.

Fig.~\ref{tu4} shows the one soliton through well with
$e^{\int{-c(t)dt}}=1-0.9 \text{sech}(t)$, while Fig.~\ref{tu5} shows
the one soliton through barrier with $e^{\int{-c(t)dt}}=1+0.9
\text{sech}(t)$. In Fig.~\ref{tu6}, the soliton passes through
multiple well or barrier with $e^{\int{-c(t)dt}}=1+h_{1}
\text{sech}(t+t_{1})+h_{2} \text{sech}(t+t_{2})$. Thereinto, $h_{1}$
and $h_{2}$ denote the height of the barrier/well, $t_{1}$ and
$t_{2}$ denote the position, and $\mid t_{1}-t_{2}\mid$ denotes the
separation distance of the barrier/well.
\\[\intextsep]
\begin{minipage}{\textwidth}
\renewcommand{\captionfont}{\scriptsize}
\renewcommand{\captionlabelfont}{\scriptsize}
\renewcommand{\captionlabeldelim}{.\,}
\renewcommand{\figurename}{Fig.\,}
\hspace{2.3cm}\includegraphics[scale=0.55]{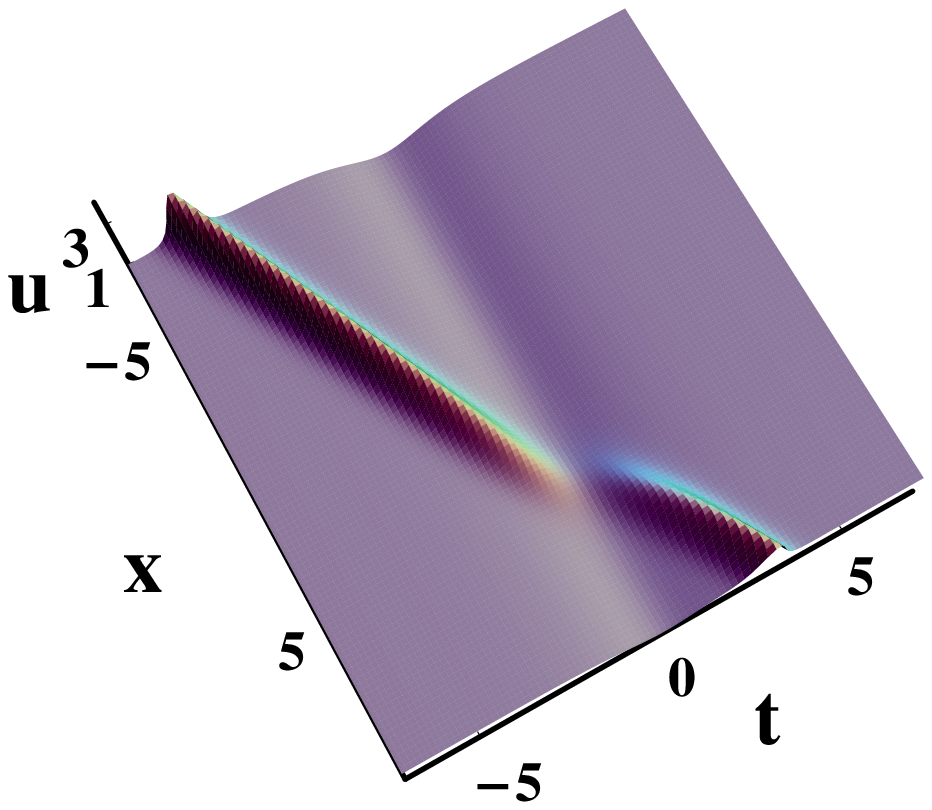}
\hspace{1.5cm}\includegraphics[scale=0.55]{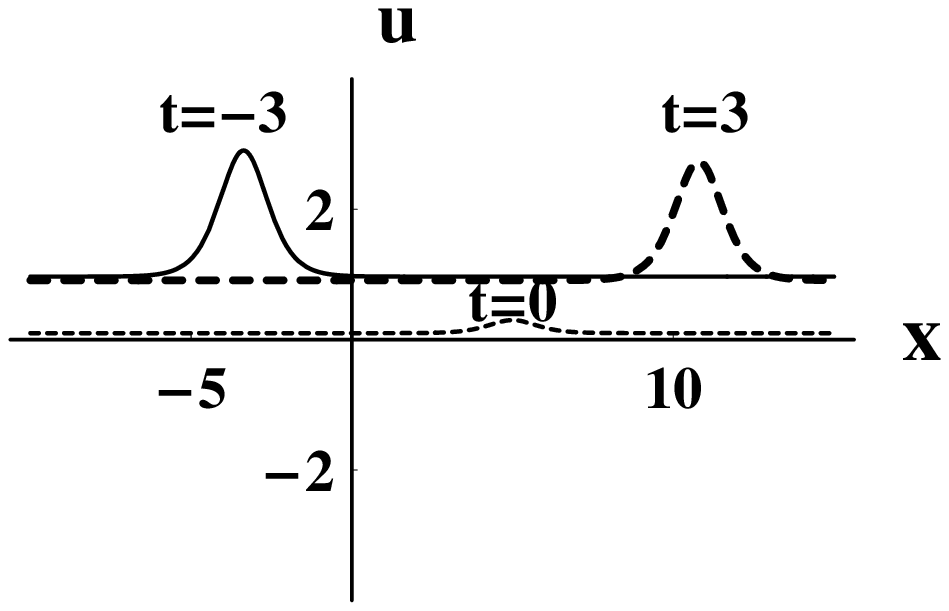}
\vspace{-0.5cm}{\center\hspace{4.7cm}\footnotesize ($a$)
\hspace{6.6cm}($b$)} \figcaption{One soliton given by
Expression~(\ref{1soliton}) with parameters: $k_{\text{1}}=2$,
$\rho=\delta=1$, $d(t)=a(t)=1$, $\xi_{\text{10}}=-10$, $f_2(t)=0$,
$e^{\int{[-a(t)\gamma(t)]dt}}=1$, $e^{\int{-c(t)dt}}=1-0.9
\text{sech}(t)$; ($b$) Profile of Fig.4 (a)  at $t=-3$, $t=0$,
$t=3$.} \label{tu4}
\end{minipage}
\\[\intextsep]
\\[\intextsep]
\begin{minipage}{\textwidth}
\renewcommand{\captionfont}{\scriptsize}
\renewcommand{\captionlabelfont}{\scriptsize}
\renewcommand{\captionlabeldelim}{.\,}
\renewcommand{\figurename}{Fig.\,}
\hspace{2.3cm}\includegraphics[scale=0.55]{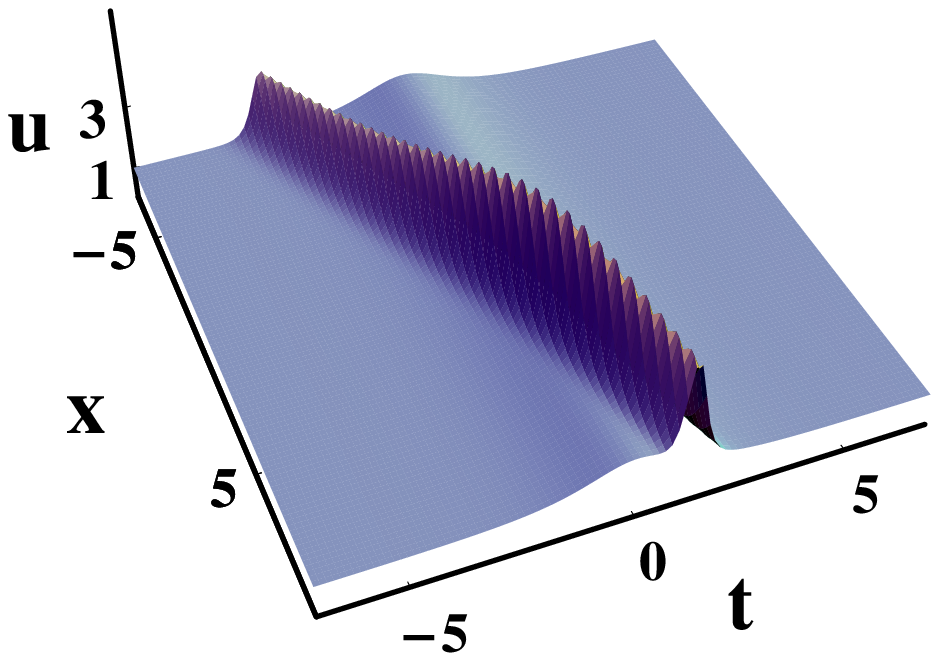}
\hspace{1.5cm}\includegraphics[scale=0.55]{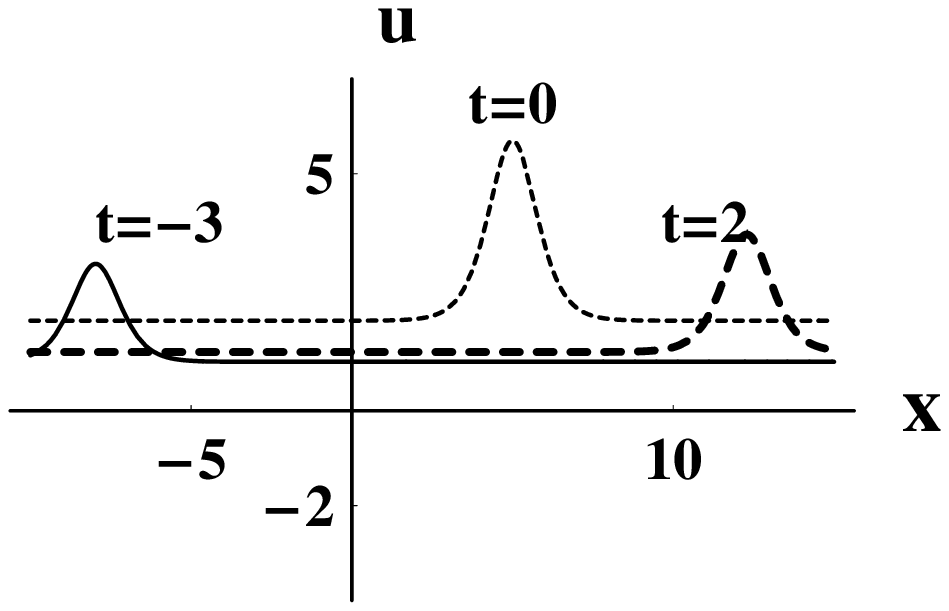}
\vspace{-0.5cm}{\center\hspace{4.7cm}\footnotesize ($a$)
\hspace{6.6cm}($b$)} \figcaption{One soliton given by
Expression~(\ref{1soliton}) with parameters: $k_{\text{1}}=2$,
$\rho=\delta=1$, $d(t)=a(t)=1$, $\xi_{\text{10}}=-10$, $f_2(t)=0$,
$e^{\int{[-a(t)\gamma(t)]dt}}=1$, $e^{\int{-c(t)dt}}=1+0.9
\text{sech}(t)$; ($b$) Profile of Fig.5 (a)  at $t=-3$, $t=0$,
$t=2$.} \label{tu5}
\end{minipage}
\\[\intextsep]
\\[\intextsep]
\begin{minipage}{\textwidth}
\renewcommand{\captionfont}{\scriptsize}
\renewcommand{\captionlabelfont}{\scriptsize}
\renewcommand{\captionlabeldelim}{.\,}
\renewcommand{\figurename}{Fig.\,}
\hspace{2.3cm}\includegraphics[scale=0.55]{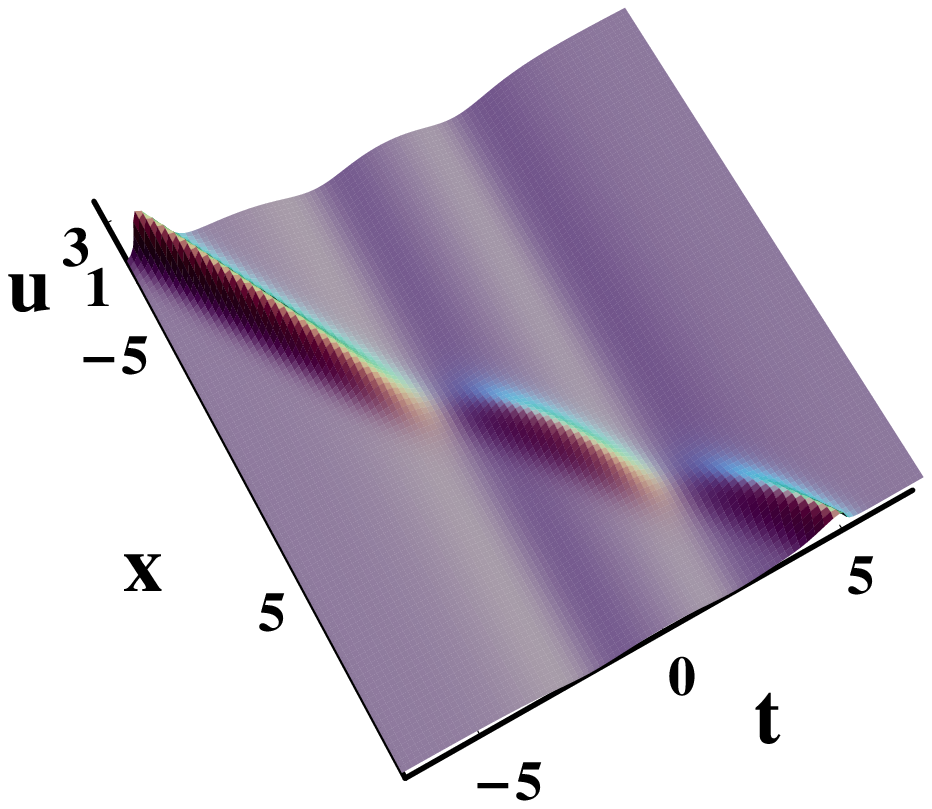}
\hspace{1.5cm}\includegraphics[scale=0.55]{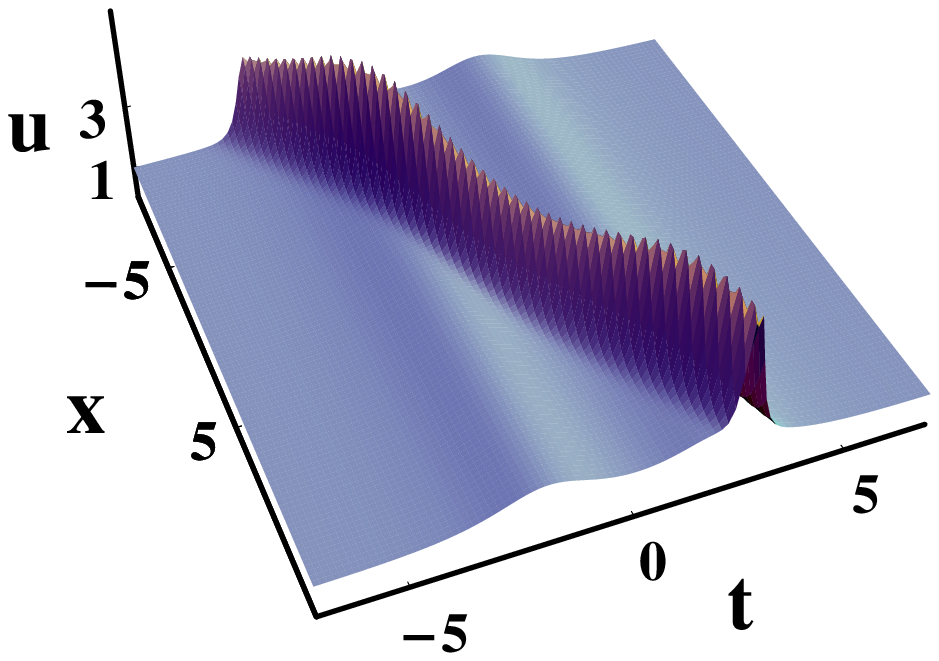}
\vspace{-0.5cm}{\center\hspace{4.7cm}\footnotesize ($a$)
\hspace{6.6cm}($b$)} \figcaption{One soliton given by
Expression~(\ref{1soliton}) with parameters: $k_{\text{1}}=2$,
$\rho=\delta=1$, $d(t)=a(t)=1$, $\xi_{\text{10}}=-10$, $f_2(t)=0$,
$e^{\int{[-a(t)\gamma(t)]dt}}=1$; ($a$) $e^{\int{-c(t)dt}}=1-0.9
\text{sech}(t-2)-0.9 \text{sech}(t+2)$  ; ($b$)
$e^{\int{-c(t)dt}}=1+0.9 \text{sech}(t-2)+0.9 \text{sech}(t+2)$.}
\label{tu6}
\end{minipage}
\\[\intextsep]

Fig.~\ref{tu7} presents the one soliton through well with periodic
background and characteristic line, and Fig.~\ref{tu8} corresponds
to the two soliton cases of Fig.~\ref{tu7}.
\\[\intextsep]
\begin{minipage}{\textwidth}
\renewcommand{\captionfont}{\scriptsize}
\renewcommand{\captionlabelfont}{\scriptsize}
\renewcommand{\captionlabeldelim}{.\,}
\renewcommand{\figurename}{Fig.\,}
\hspace{2.3cm}\includegraphics[scale=0.55]{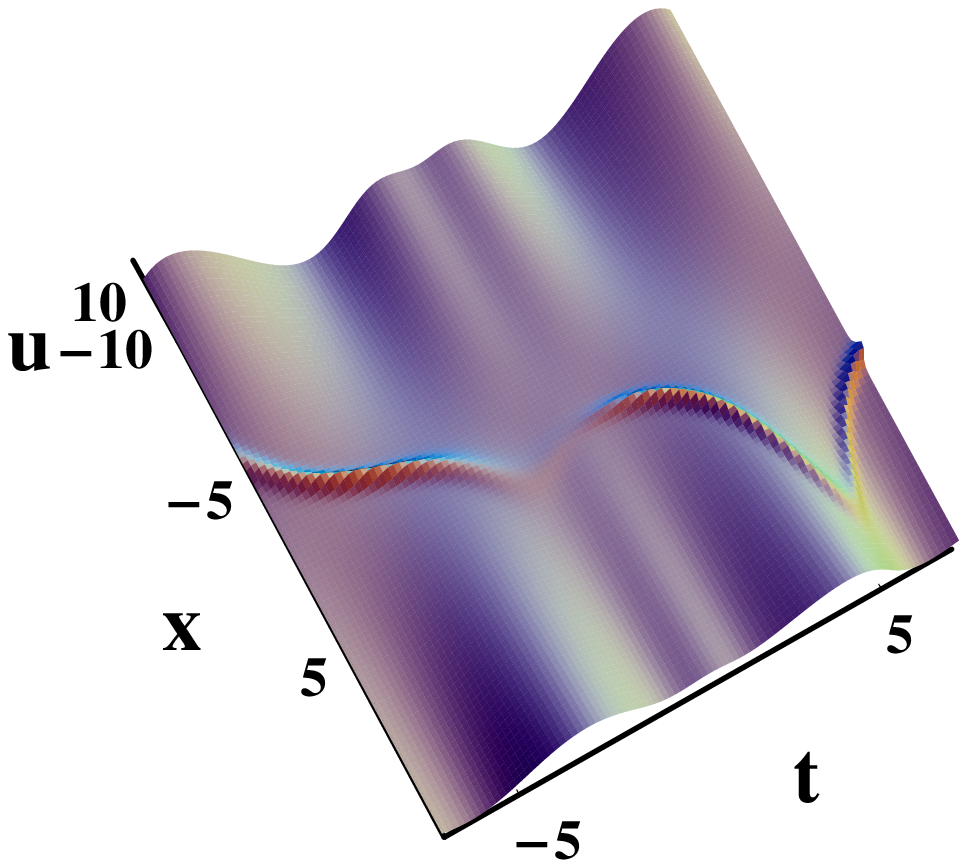}
\hspace{1.5cm}\includegraphics[scale=0.55]{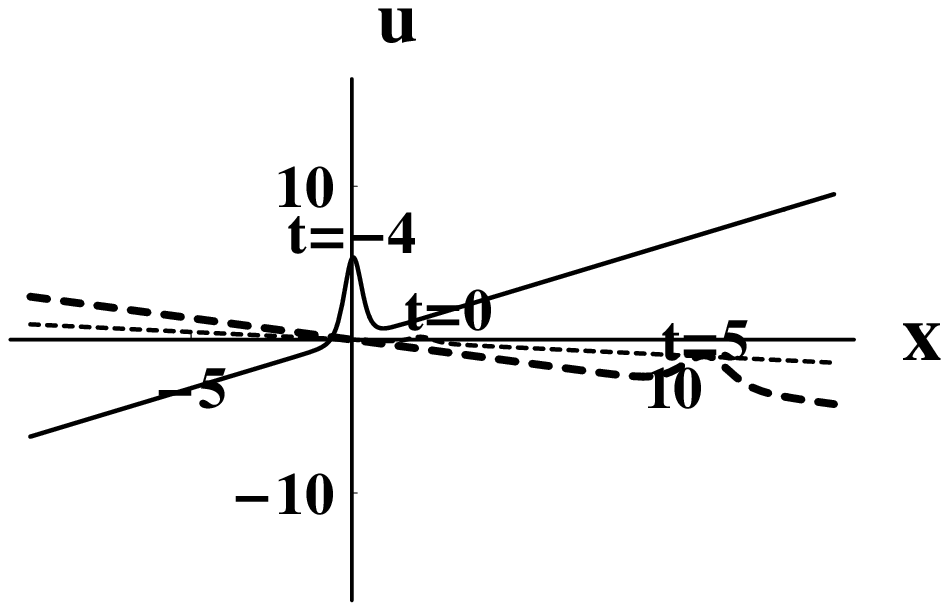}
\vspace{-0.5cm}{\center\hspace{4.7cm}\footnotesize ($a$)
\hspace{6.6cm}($b$)} \figcaption{One soliton given by
Expression~(\ref{1soliton}) with parameters:$k_{\text{1}}=2$,
 $\rho=1$, $d(t)=f_{2}(t)=\delta=0$,
$\xi_{\text{10}}=-10$, $a(t)=\{[2+\sin(t)][1-0.9
\text{sech}(t)]\}^{-1}$, $e^{\int{[-a(t)\gamma(t)]dt}}=2+\sin(t)$,
$e^{\int{-c(t)dt}}=1-0.9 \text{sech}(t)$; ($b$) Profile of Fig.7 (a)
at $t=-4$, $t=0$, $t=5$.} \label{tu7}
\end{minipage}
\\[\intextsep]
\\[\intextsep]
\begin{minipage}{\textwidth}
\renewcommand{\captionfont}{\scriptsize}
\renewcommand{\captionlabelfont}{\scriptsize}
\renewcommand{\captionlabeldelim}{.\,}
\renewcommand{\figurename}{Fig.\,}
\hspace{2.3cm}\includegraphics[scale=0.55]{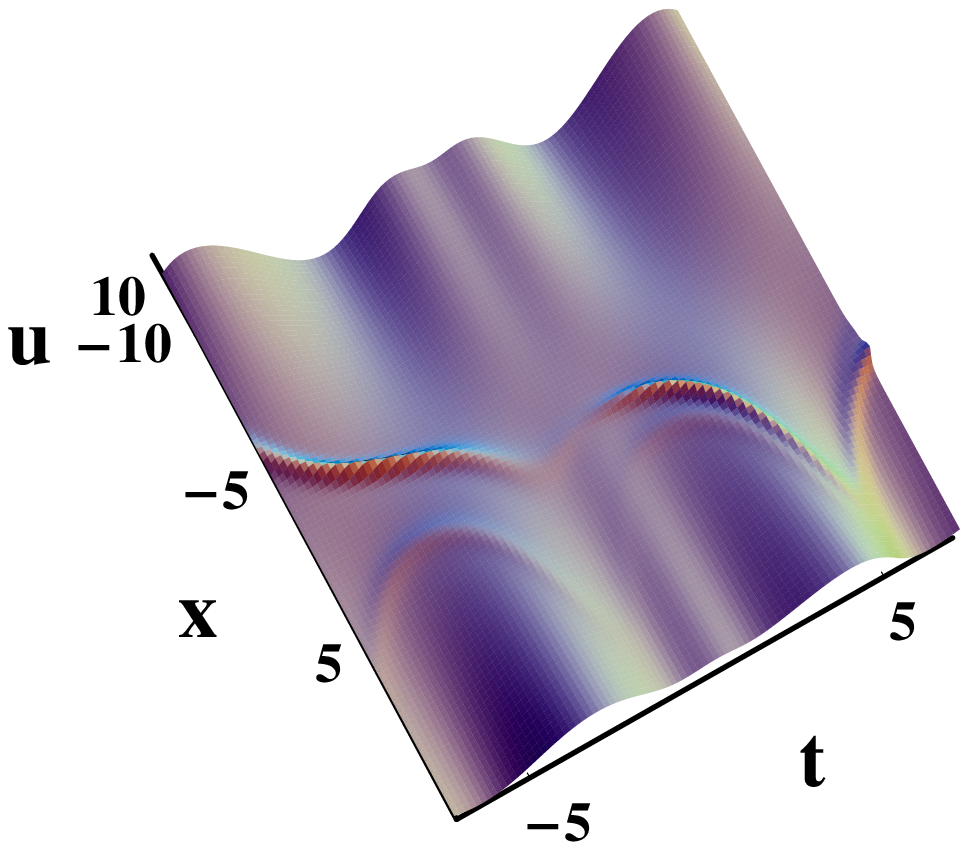}
\hspace{1.5cm}\includegraphics[scale=0.55]{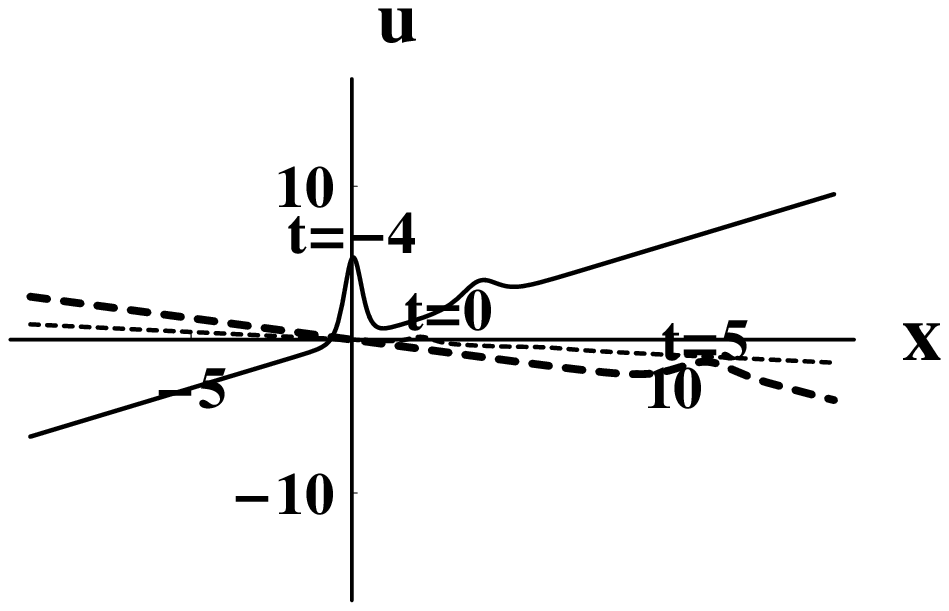}
\vspace{-0.5cm}{\center\hspace{4.7cm}\footnotesize ($a$)
\hspace{6.6cm}($b$)} \figcaption{Two solitons given by
Expression~(\ref{2soliton}) with parameters: $k_{\text{1}}=2$,
$k_{\text{2}}=1$, $\rho=1$, $d(t)=f_{2}(t)=\delta=0$,
$\xi_{\text{10}}=-10$, $a(t)=\{[2+\sin(t)][1-0.9
\text{sech}(t)]\}^{-1}$, $e^{\int{[-a(t)\gamma(t)]dt}}=2+\sin(t)$,
$e^{\int{-c(t)dt}}=1-0.9 \text{sech}(t)$; ($b$) Profile of Fig.8 (a)
at $t=-4$, $t=0$, $t=5$.} \label{tu8}
\end{minipage}
\\[\intextsep]

\vspace{7mm} \noindent {\Large{\bf V. Conclusions}}

\vspace{3mm}In this paper, Eq.~(\ref{equation}), a
variable-coefficient model with spacial inhomogeneity in
fluids~\cite{hd,tkg,hd3,gxh,hhd,ph,hd2,rhjg,gae,tangxiaoyan},
 is investigated with symbolic computation.
Under coefficient constraints~(\ref{cc1}) and~(\ref{cc2}),
Eq.~(\ref{equation}) is transformed into its bilinear form, and the
multi-soliton solutions are constructed.  The function $\gamma(t)$
corresponds to spacial inhomogeneity, and the nonlinear coefficient
$a(t)$ can also affect the soliton width and amplitude for the
existence of $\gamma(t)$, as shown in  Figs.~\ref{tu1}-~\ref{tu3}.

Nonlinear tunneling for  Eq.~(\ref{equation}) is a special state of
amplitude, so it can be regarded as a kind of variable coefficient
effects. With $e^{\int{-c(t)dt}}$ taken as $1+\sum h_{n}
\text{sech}(t+t_{n})$, nonlinear tunneling in
Figs.~\ref{tu4}-~\ref{tu6} is illustrated, where $h_{n}$  denotes
the height of the barrier/well, $t_{n}$ denotes the position, and
$\mid t_{n}-t_{l}\mid$ denotes the separation distance of the
barrier/well. Finally, Figs.~\ref{tu7} and~\ref{tu8} display the
combination of nonlinear tunneling and variable coefficient effects.

\vspace{7mm} \noindent {\Large{\bf Acknowledgements}}

\vspace{3mm}We express our sincere thanks to all the  members of our
discussion group for their valuable comments. This work has been
supported by the National Natural Science Foundation of China under
Grant No. 11302014, and by the Fundamental Research Funds for the
Central Universities under Grant Nos. 50100002013105026 and
50100002015105032 (Beijing University of Aeronautics and
Astronautics).

\end{document}